\documentclass[twocolumn,prl,showpacs,preprintnumbers,amsmath,amssymb]{revtex4}
\usepackage{graphicx}
\usepackage{color}
\usepackage{setspace}
\graphicspath{{./figures/}}
\newcommand{\futomoji}[1]{\mbox{\boldmath$#1$}}

\begin{document}
\title{Supersolid states in a spin system 
--- phase diagram and collective excitations}
\author{Yuta Murakami$^{1}$, Takashi Oka$^2$, Hideo Aoki$^1$ }
\affiliation{$^1$Department of Physics, The University of Tokyo,
Hongo, Tokyo 113-0033, Japan\\
$^2$ Department of Applied Physics, The University of Tokyo, Hongo, Tokyo 113-8656, Japan}
\date{\today}

\begin{abstract}
Phases analogous to supersolids can be realized in spin systems. 
Here we obtain the phase diagram of a frustrated dimer spin-1/2 system on a square lattice and study the collective excitation spectra, 
focusing on the supersolid state (SS). 
In the phase diagram on a parameter space of the exchange interaction and magnetic field,  
we find, on top of the SS phase, a phase that has no counterpart in the Bose Hubbard model and this state becomes dominant at the region where the enhancement of SS occurs in the bose Hubbard model. 
We then investigate the excitation spectrum
and spin-spin correlation, which can be detected by neutron scattering experiments. We obtain an analytic expression for the spin wave velocity, 
which agrees with hydrodynamic relations. The intensity of excitation modes in the spin-spin correlation function is calculated and their change
in the supersolid and superfluid states is discussed.

\end{abstract}

\maketitle
\setstretch{0.9}
\section{Introduction}
Supersolid (SS) state is a phase where both off-diagonal long-range order and diagonal-long range order coexist. After the non-classical rotational inertia experiment in $\mathrm{He}^4$ suggested an SS phase, the SS state attracted considerable interest\cite{He4}. However, the interpretation of the result is still controversial\cite{He4_2}, 
and efforts to find SS continue. 
Lattice systems are other candidates to find SS than $\mathrm{He}^4$.
 Cold atoms on optical lattices are one of them. 
A proposal \cite{coldextend} was made based on the 
fact that the extended bose Hubbard model with a nearest neighbor interaction shows SS\cite{Hubbardss1,Hubbardss2}. Dipole-dipole interaction is also suggested to help realizing SS\cite{polerss}.

\begin{figure}[h]
  \begin{center}
   \includegraphics[width=80mm]{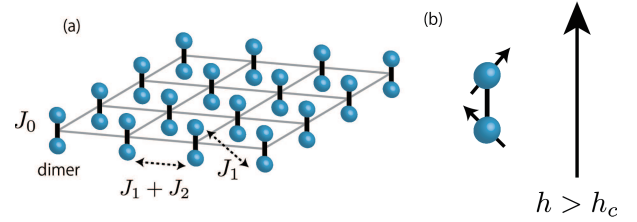}
   \caption{(a) The frustrated spin-1/2 dimer Heisenberg  model considered 
here. (b) Spin configuration in a dimer is schematically 
shown when the magnon BEC occurs 
for $h \equiv g\mu _B H$ exceeding a critical value($h_c$).}
   \label{Fig:ourmodel}
  \end{center}  
\end{figure}

Another, entirely different avenue to find SS in lattice systems is to 
consider quantum magnets. This is because
certain spin systems can be effectively regarded as bose systems \cite{magnonBEC1,magnonBEC2,magnonBEC3}. After a theoretical proposal of SS in a dimer spin system \cite{SSspin1}, spin systems are attracting much attention as a promising candidate to find SS phases \cite{SSspin1,SSspin2,SSspin3,SSspinisotropy,SSspinisotropy2,SSspinisotropy3}. 
One example is the spin-1 Heisenberg model with an anisotropy, which is effectively obtained from a spin-1 frustrated dimer model \cite{SSspin2}.  
Another example is the spin-1/2 dimer model
with large Ising-like exchange anisotropy \cite{SSspin1,SSspin3}.
The anisotropy can be effectively realized in a lattice with large frustration.  Ref.\cite{SSspinisotropy} has 
investigated a frustrated spin-1/2 spin-dimer Heisenberg model with spin-isotropic couplings on square lattice(Fig.\ref{Fig:ourmodel}), and shown that 
the model actually exhibits a SS state.
However, the study was concentrated on a specific choice of 
parameters, while the phase diagram in a wider parameter space has yet to be determined.  
In addition, experimentally relevant properties of the SS phase  
have not been fully understood, either.  

These have motivated us, in the present work, to investigate the phase diagram 
and dynamical properties of the frustrated spin-1/2 dimer Heisenberg model on square lattice.  
There, we employ the bond operator method as well as the generalized spin wave theory. 
In the phase diagram on a parameter space of the exchange interaction and 
the external magnetic field,  we have found, on top of the SS phase, 
a phase that has no counterpart in the bose Hubbard model. 
As for the dynamical properties, we study the excitation spectrum as seen in spin-spin correlations to provide important information that is
 measurable with inelastic neutron scattering.
In the language of the cold atoms on optical lattices, one of the correlation 
functions studied here is equivalent to the dynamical structure factor. 
We have revealed the behavior of the dynamical properties in the SS and SF phases, especially around their phase boundaries,
 which can be used as a probe to detect the phase transition experimentally. 
We have also obtained an analytic expression for the spin wave velocity, 
which agrees with hydrodynamic relations. 

\section{Formalism}
The Hamiltonian of the frustrated spin-1/2 dimer Heisenberg  model (Fig.\ref{Fig:ourmodel}) is
\begin{equation}
{\scriptstyle
\begin{split}
{\cal H} = &J_0 \displaystyle \sum_{i}\futomoji {S}_{1i}\cdot \futomoji {S}_{2i} +J_1 \displaystyle \sum_ {\langle i,j\rangle}\futomoji{S}_{i} \cdot \futomoji{S}_{j}\\&+J_2\displaystyle \sum_ {\langle m,i,j\rangle}  \futomoji{S}_{mi} \cdot \futomoji{S}_{mj}
-g\mu _B H \displaystyle \sum_ {i}  S^z_{i}, 
\label{eq:ourmodel}
\end{split}}
\end{equation}
where $i,j$ label the dimer's central position, $m=1,2$ labels
the sites within each dimer, and $\futomoji{S}_{i}\equiv \futomoji{S}_{1i}+\futomoji{S}_{2i}$.  Antiferromagnetic exchange interactions comprise 
intra-dimer ($J_0$), inter-dimer ($J_1, J_2$) interactions with 
$J_0,J_1,J_2>0$.  Frustration is caused by the $J_1$ coupling between inequivalent sites between adjacent dimers.  
Hereafter we take $J_0=1$ as a unit of energy.  
In the Zeeman energy term,  
$g$ is the g-factor, $\mu_B$ the Bohr magneton, 
and $H$ the external magnetic field.

For treating the spin system, we can adopt 
the bond operator formalism \cite{bondtlcucl3,bondmethod2,bondop1}, in which 
the four spin states (one singlet and three triplet) 
in a dimer are described by 
four bosonic operators, $s$ and $t$, as
\begin{equation}
{\scriptstyle
\begin{split}
 & t_{+}^{\dagger }|{\rm vac}\rangle=-|1,1\rangle ,\\
s^{\dagger}|{\rm vac}\rangle=|0,0\rangle,\:\:& t_{0}^{\dagger }|{\rm vac}\rangle=|1,0\rangle, \\ 
 &t_{-}^{\dagger }|{\rm vac}\rangle=|1,-1\rangle.
 \end{split}}
 \label{eq:bondop}
\end{equation}
Here $|0,0\rangle$ and $|1,\alpha\rangle$ ($\alpha = \pm,0$) 
stand for singlet and triplet states, respectively, 
while $|{\rm vac}\rangle$ denotes the vacuum of $s,t$.  
The transformation becomes exact when a constraint, 
$s^{\dagger}s+\sum_{\alpha=\pm,0 }t^{\dagger}_{\alpha}t_{\alpha}=1$, 
is imposed.
With the bond operators, the Hamiltonian (\ref{eq:ourmodel}) is expressed as 
\begin{equation}
{\scriptstyle
\begin{split}
H=&-\frac{3}{4}J_0\sum_is_{i}^{\dagger}s_{i}+\left(\frac{J_0}{4}-h\right)\sum_i t_{+i}^{\dagger} t_{+i}\\&
+\frac{J_0}{4}\sum_i t_{0i}^{\dagger}t_{0i} +\left(\frac{J_0}{4}+h\right)\sum_i t_{-i}^{\dagger}t_{-i}
\\ &+J_2/2\sum_{\langle i,j\rangle}H_{st}(i,j) + (J_1+J_2/2)\sum_{\langle i,j\rangle}H_{tt}(i,j).
\label{eq:bondrep}
\end{split}}
\end{equation}
Here $h \equiv g\mu _B H$, and
\begin{equation}
H_{st}(i,j)=\sum_{\alpha=\pm,0}(t_{i\alpha }^{\dagger}t_{j\alpha}s^{\dagger}_js_i+t_{i\alpha }^{\dagger}t^{\dagger}_{j\bar{\alpha}}s_js_i+{\rm H.c.}),
\end{equation}
\begin{equation}
{\scriptstyle
\begin{split}
H_{tt}(i,j) = &[t^{\dagger}_{j0}t_{i0}(t^{\dagger}_{i+}t_{j+}+t^{\dagger}_{i-}t_{j-})+{\rm H.c.}]\\
             &-[{t_{i0}t_{j0}(t^{\dagger}_{i+}t^{\dagger}_{j-}+t^{\dagger}_{i-}t^{\dagger}_{j+})+{\rm H.c.}}]\\
             &+(t^{\dagger}_{i+}t_{i+}-t^{\dagger}_{i-}t_{i-})(t^{\dagger}_{j+}t_{j+}-t^{\dagger}_{j-}t_{j-}),\\
\end{split}}         
\end{equation}
where $\bar{\alpha}=\mp,0$ for $\alpha=\pm,0$ respectively.
Using this expression, we can obtain a variational 
ground-state wavefunction,
\begin{equation}
{\scriptstyle
\begin{split}
|{\rm GS}\rangle &= \prod_{i\in A} (y_As_i^{\dagger}+\sum_{\alpha}x_{A\alpha}t_{i\alpha}^{\dagger})\\
&\times \prod_{i\in B} (y_Bs_i^{\dagger}+\sum_{\alpha}x_{B\alpha}t_{i\alpha}^{\dagger})|{\rm vac}\rangle,
\label{eq:bondgs}
\end{split}}
\end{equation}
where the coefficients $x, \; y$ are complex in general, 
and determined numerically. 
Here we have divided the square lattice into checkerboard sublattices (A and B), and 
since difference between sublattices is allowed, this wave function is
capable of describing a SS state.

A simplification occurs when the $|1,0\rangle$ state is ignored.
We have numerically confirmed that this is in fact permissible 
in most parts of the phase diagram($x_0=0$.).  
Then the Hamiltonian becomes
\begin{equation}
{\scriptstyle
\begin{split}
H_{\rm eff}=\frac{J_2}{2}\sum_{\langle i,j\rangle}
[(t_{i+}^{\dagger}s_{i}+s_{i}^{\dagger}t_{i-})(s_{j}^{\dagger}t_{j+}+t_{j-}^{\dagger}s_{j})+{\rm H.c.}]\\
+\left(J_1+\frac{J_2}{2}\right)\sum_{\langle i,j \rangle}(t_{i+}^{\dagger}t_{i+}-t_{i-}^{\dagger}t_{i-})(t_{j+}^{\dagger}t_{j+}-t_{j-}^{\dagger}t_{j-})\\ 
+J_0\sum_{i}(t_{i+}^{\dagger}t_{i+}+t_{i-}^{\dagger}t_{i-})
-h\sum_i(t_{i+}^{\dagger}t_{i+}-t_{i-}^{\dagger}t_{i-})
\label{eq:effective1} 
\end{split}
}
\end{equation}
with a constraint $s_i^{\dagger}s_i+\sum_{\alpha=\pm}t^{\dagger}_{i\alpha}t_{i\alpha}=1$. 
This coincides with the anisotropic spin-1 Heisenberg model used in  Ref.~\cite{SSspin2}.

We can effectively identify this model with an extended bose 
Hubbard model,
\begin{eqnarray}
H=-t\sum_{\langle i,j\rangle}(a^{\dagger}_ia_j+{\rm H.c.})-\mu\sum_in_i \\ 
+V\sum_{\langle i,j\rangle}n_in_j+U\sum_in_in_i,\label{eq:extendH}
\end{eqnarray}
where $a^{\dagger}$ is the boson creation operator, $\langle i,j\rangle$ nearest neighbors, $\mu$ the boson chemical potential, $U$ the on-site Hubbard 
interaction and $V$ the nearest-neighbor interaction.  
To do this, we  truncate the states  to three states up to the doubly-occupied state in the extended Hubbard model and identify $t^{\dagger}_{-}|{\rm vac}\rangle=|0\rangle,\;
s^{\dagger}|{\rm vac}\rangle=|1\rangle,\;\;t^{\dagger}_{+}|{\rm vac}=\rangle|2\rangle$, and 
express Eq.(\ref{eq:extendH}) in terms of $t,\;s$ as in Ref.\cite{Hubbarddynam1}. This procedure is known as the Schwinger-boson approach.
Then it turns out that the spin model Eq. (\ref{eq:effective1}) can be regarded as the extended Hubbard model through 
\begin{equation}
J_2/2\leftrightarrow -t, \;\;\;J_1+J_2/2\leftrightarrow V,\;\;\;J_0\leftrightarrow U/2, \;\;\; h\leftrightarrow \delta\mu,
\label{eq:corresponde}
\end{equation}
where $\delta\mu \equiv \mu-\frac{U}{2}-ZV$ with 
$Z$ being the coordination number. 
Strictly speaking, there is only slight difference in the hopping terms.  
In the truncated Hubbard model it is written as 
$-t\sum_{\langle i,j\rangle}[(\sqrt{2}t_{i+}^{\dagger}s_{i}+s_{i}^{\dagger}t_{i-})(\sqrt{2}s_{j}^{\dagger}t_{j+}+t_{j-}^{\dagger}s_{j})+{\rm H.c.}]$, where $\sqrt{2}t_{i+}^{\dagger}s_{i}+s_{i}^{\dagger}t_{i-}$ corresponds to 
the creation operator $a^{\dagger}$ in the truncated space. On the other hand, in the case of the spin model, we regard $t_{i+}^{\dagger}s_{i}+s_{i}^{\dagger}t_{i-}$ 
as the creation of a boson, see eq.(\ref{eq:effective1})\cite{magnonBEC4}.

Thus the spin-1/2 model is effectively a 
{\it semi-hard core boson system} in 
regions in the phase diagram where $|1,0\rangle$ state can be ignored\cite{magnonBEC4}. 
In order to compare the phase diagrams of the two systems,
it is useful to draw the phase diagrams with the parameter corresponding to $V$ fixed, since phase diagrams of the extended bose Hubbard model is often written in this way. 

 \begin{center}
    \begin{table*}
  \begin{tabular}{|l|l|}
  \hline
  Spin-1/2 dimer system&Bose system\\
  \hline
States: $t^{\dagger}_{-}|{\rm vac}\rangle,\;s^{\dagger}|{\rm vac}\rangle,\;t^{\dagger}_{+}|{\rm vac}\rangle$&States: $|0\rangle\;,|1\rangle\;,|2\rangle$\\
  \hline
In-plane magnetization: $M_{xy}=\langle S^+_{1}-S^+_{2} \rangle /\sqrt{2}$&Order parameter: $\langle a^{\dagger}\rangle$\\
  \hline
 In-plane staggered magnetization: & Superfluid density: \\$n_s=(\frac{1}{N}\sum_i\langle S^+_{i,1}-S^+_{i,2}\rangle e^{i\futomoji{Q}\cdot\futomoji{r}_i}/\sqrt{2})^2$ & $n_s=|\frac{1}{N}\sum_i \langle a_i^{\dagger}\rangle|^2$ \\
  \hline
  Staggered magnetization: &Staggered occupation number:\\$m_{z}^{\rm st}=\frac{1}{2N}\sum_i\langle S^z_{i,1}+S^z_{i,2} \rangle e^{i\futomoji{Q}\cdot\futomoji{r}_i}$&$n^{\rm st}=(n_A-n_B)/2$ \\
  \hline
Averaged magnetization: $m_z=\frac{1}{2N}\sum_i\langle S^z_{i,1}+S^z_{i,2}\rangle$&Averaged occupation number: $n=(n_A+n_B)/2$ \\
  \hline
  \end{tabular}
  \caption{Correspondence between the dimer-spin system and the bose system. 
In the bose system, $|n\rangle$ represents a state with $n$ bosons, and $n_A, n_B$ is the boson density on A, B sublattices, respectively.  For the spin system the bond-operator representation is used.} 
  \end{table*}
  \end{center}

    \begin{table}
  \begin{tabular}{|c|l|}
  \hline
  Spin-1/2 dimer system&Bose system\\
  \hline
  $n_s=0,m_z^{\rm st}=0,$& Mott-insulating phase (MI)   \\
   $m_z=0.0$& for $n=1$\\
    \hline
  $n_s=0,m_z^{\rm st}=0$& Mott-insulating phase (MI)  \\
    $m_z=0.5$& for $n=2$\\
  \hline
    $n_s\neq0, m_z^{\rm st}=0$& Superfluid (SF) \\
  \hline
    $n_s=0,m_z^{st}\neq0$& Charge ordered state (CO) \\
     \hline
    $n_s\neq0,m_z^{\rm st}\neq0$& Supersolid (SS) \\
  \hline
  \end{tabular}
  \caption{Correspondence of phases between the dimer-spin system 
and the bose system. $n$ is the averaged occupation number of Boson. } 
  \end{table}

From the above correspondence, we can naturally define the order parameters: The in-plane magnetization $M_{xy,i}=\langle S^+_{i,1}-S^+_{i,2} 
\rangle /\sqrt{2}$, which denotes the difference 
of the magnetization in the spins in each dimer,
represents the breakdown of the U(1) symmetry in the spin model, which 
corresponds to $\langle a^{\dagger}\rangle$ in the bose Hubbard model 
with a broken U(1) symmetry (the superfluid density in BEC).
 Thus the averaged superfluid density is $n_s=(\frac{1}{N}\sum_i\langle S^{+}_{i,1}-S^{+}_{i,2}\rangle e^{i\futomoji{Q}\cdot\futomoji{r}_i}/\sqrt{2})^2 $, 
where $\futomoji{Q}=(\pi,\pi)$ and $N$ the total number of sites.  
Note that the factor $e^{i\futomoji{Q}\cdot\futomoji{r}_i}$ takes care 
of the fact that $J_2>0$ induces antiferroic ordering. In 
the terminology of the bose Hubbard model,  this is because the hopping term is positive, see Eq.(\ref{eq:corresponde}).
The $z$-component staggered 
magnetization, $m_{z}^{\rm st}=\frac{1}{2N}\sum_i\langle S^z_{i,1}+S^z_{i,2} \rangle e^{i\futomoji{Q}\cdot\futomoji{r}_i}$, 
represents the breaking of the $Z_2$ symmetry (the 
symmetry between A,B sublattices).
 Another important quantity is 
the uniform magnetization $m_z=\frac{1}{2N}\sum_i\langle S^{z}_{i,1}+S^{z}_{i,2}\rangle$, which corresponds to the density of bosons in 
the Hubbard model.  The correspondence is summarized in table 1.
\begin{figure}[h]
 \begin{center}
\includegraphics[width=80mm]{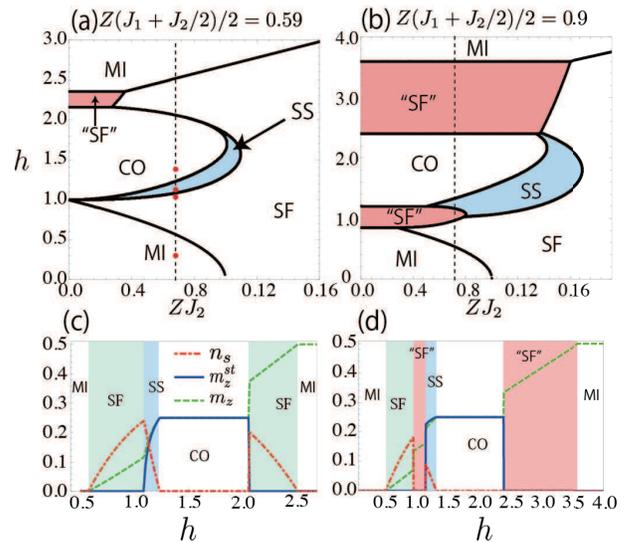}
  \caption{Ground-state phase diagram 
for $Z(J_1+J_2/2)/2=0.59$ (a), and $0.9$ (b) 
with $Z$ being the coordination number
(i.e., $ZV=0.59U$ (a), $ZV=0.9U$ (b) for the boson model).  
The parameters $h$ and $ZJ_2$ correspond to 
the chemical potential $\delta \mu$ 
and hopping parameter $-t$, respectively, in the boson system. 
Phases are denoted by 
SS (supersolid), SF (superfluid), MI (Mott insulator), and CO 
(charge-ordered phase), while ``SF" stands for a phase that 
has no counterpart in the bose system. 
Various order parameters are plotted against $h$ in 
(c) for $Z(J_1+J_2/2)/2=0.59, ZJ_2=0.68$, and in 
(d) for $Z(J_1+J_2/2)/2=0.9, ZJ_2=0.72$.  
The dashed lines in (a,b) indicate that the cross sections 
at which (c,d) are plotted, respectively. 
Red dots in (a) indicate the points at which the 
excitation spectra are displayed in Fig.\ref{Fig:excitation} 
below.
}
  \label{Fig:phase}
 \end{center}
\end{figure}

\section{Phase diagram}

Figure \ref{Fig:phase} shows the phase diagram 
against $h$ and $J_2/2$ obtained by optimize the variational wave function Eq.(\ref{eq:bondgs}).  
The plot is a counterpart in the spin system of a phase diagram for the bose Hubbard model
against the chemical potential ($\delta \mu$) and hopping parameter 
($-t$).  
The two panels (a,b) correspond to different values of $ZV$,
which is the effective nearest neighbour interaction $V=J_1+J_2/2$ multiplied by the 
coordination number $Z$.  
Thus panels (a),(b) for $Z(J_1+J_2/2)/2=0.59, 0.9$ correspond to 
those for the bose system at $ZV=0.59U, 0.9U$, respectively. 
To specify the phases, here we adopt the 
terminology from the bose Hubbard model, see table 2. In particular, 
superfluid (SF) is a phase with  $n_s\neq 0$ and $m_z^{\rm st}=0$, 
while a supersolid (SS) state is a phase with $n_s\neq0$ and 
$m_z^{\rm st}\neq0$ simultaneously. 

First, the SS phase in the present spin model 
appears adjacent to, and mainly in the lower-half of, 
the CO phase.  
We note that this feature is also seen in the phase diagram of the 
extended bose-Hubbard model\cite{Hubbardss2}. 
At the boundary of MI the excitation gap closes, and 
the boundary can be given analytically, 
where the lower density branch is $h=\sqrt{1-ZJ_2}$, while the 
higher one is  $1+ZJ_1+ZJ_2$, within our approximation.  
Figure \ref{Fig:phase}(c) shows the $h$ dependence 
(on a cross section indicated in (a)) 
of relevant order parameters for $ZJ_1=0.84$ and $ZJ_2=0.68$.
The results agree qualitatively
with those obtained with the infinite time-evolving block decimation 
(iTEBD) combined with the tensor renormalization-group
(TRG) approach\cite{SSspinisotropy}. 
This supports the validity of our method for this model.

Interestingly, we find regions where we cannot neglect the 
existence of $t^\dagger_0$ (we call this phase as ``SF"), 
while in other regions we can.  In the ``SF" regions 
(red regions in Fig.\ref{Fig:phase} (a,b)), 
the value of the coefficient of $s^{\dagger}$ in Eq.(\ref{eq:bondgs}) is 0, 
while U(1) symmetry is broken. The wave function takes a form 
$|{\rm GS}\rangle=\prod_{i\in A} (x_{+}t_{i+}^{\dagger}+x_0t_{i0}^{\dagger}+x_{-}t_{i-}^{\dagger})\times \prod_{i\in B} (-x_{+}t_{i+}^{\dagger}+x_{0}t_{i0}^{\dagger}-x_{-}t_{i-}^{\dagger})|{\rm vac}\rangle$, where $x_{\alpha}$ is real.  
This state may be thought of as a canted antiferromagnetic 
state, which can appear in the simplest isotropic spin-1 Heisenberg model in an external magnetic field.
Comparing Fig.\ref{Fig:phase} (a) and (b),
we notice that the SS region becomes wider as the repulsion becomes
stronger. The ``SF" region also expands, where SS and ``SF" 
phases compete with each other. 
Fig.\ref{Fig:phase}(d) plots relevant order parameters 
against $h$ (on a 
cross section indicated in (b)) for $ZJ_1=1.44$ and $ZJ_2=0.72$. 
For this set of parameters, phase transitions occur six times as 
the external field is increased.  Specifically, 
the transition from SF to ``SF" is seen to be discontinuous. 

Let us compare the present result with that for 
the extended bose Hubbard model.  In the latter, 
SS region becomes wider when the nearest-neighbor repulsion $V$ 
is increased. Moreover, when $ZV>U$, there is no MI and all insulating phases
are CO, while the SS region becomes even wider \cite{SSenhance}. 
These behaviors are contrasted with the present 
phase diagram for the spin model for $ZV>U$, 
where, contrary to a naive expectation, it turns out that SS does 
not expand, but gives way to ``SF" for a fixed $V$, and that 
the SS region is completely suppressed by ``SF" for large enough $V$ (not shown).  Thus we do have differences between the spin and bose models.

\section{Excitations}
Let us move on to the study of excitation spectra. For this 
we employ a generalized spin wave theory which is applicable except for the ``SF" phase\cite{bondop1}.  
We convert the Hamiltonian Eq.(\ref{eq:bondrep}) 
into an effective one by introducing boson operators 
$\{b\}$ with a canonical transformation, 
\begin{equation}
\begin{split}
&b^{\lambda}_{0i}=u_{\lambda}s_i+v_{\lambda}(f_{\lambda}t_{+i}+g_{\lambda}t_{-i}),\\
&b^{\lambda}_{+i}=-v_{\lambda}s_i+u_{\lambda}(f_{\lambda}t_{+i}+g_{\lambda}t_{i-}),\\
&b^{\lambda}_{00i}=t_{0i},\\
&b^{\lambda}_{-i}=-g_{\lambda}t_{i+}+f_{\lambda}t_{i-}.
\end{split}\label{eq:unitrans}
\end{equation}
Here, $\lambda =$ A or B, $u$,$v$,$f$ and $g$ 
(with $u^2+v^2=1$ and $f^2+g^2=1$) are real and defined 
in such a way that the ground state is $(\prod_{i\in A} b^{A\dagger}_{0i})(\prod_{j\in B} b^{B\dagger}_{0j})|{\rm vac}\rangle$.  After this transformation, the constraint is converted to 
$b^{\lambda\dagger}_{0i}b^{\lambda}_{0i}+\sum_{\theta}b^{\lambda\dagger}_{\theta,i}b^{\lambda}_{\theta,i}=1$, where $\theta=\pm,00$.  
Then we deal with the constraint in terms of the Holstein-Primakoff (HP) transformation,
\begin{equation}
b^{\lambda}_{0i}=b^{\lambda\dagger}_{0i} = \left(1-\sum_{\theta}\hat{n}^{\lambda}_{\theta,i}\right)^{1/2},
\end{equation}
where $\hat{n}^{\lambda}_{\theta,i}=b^{\lambda\dagger}_{\theta,i}b^{\lambda}_{\theta,i}$.  

\begin{figure}[h]
  \begin{center}
   \includegraphics[width=90mm]{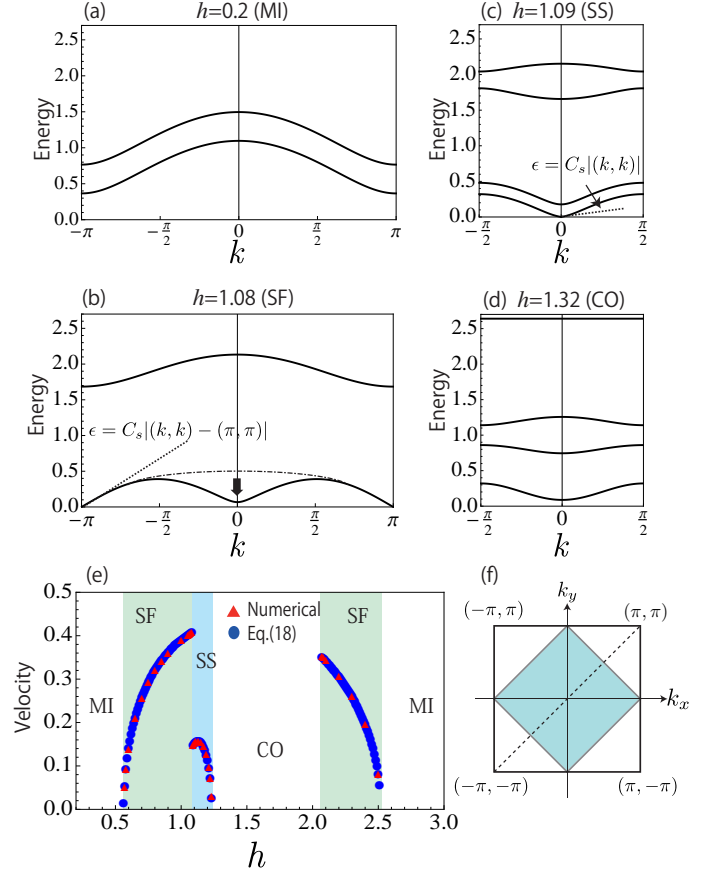}
  \end{center}
  \caption{Excitation spectra against $k_x=k_y=k$ 
for a Mott insulator at 
$h=0.2$ (a), superfluid at $h=1.08$ (b), 
supersolid at $h=1.09$ (c), and charge order at $h=1.32$ (d), with 
$ZJ_1=0.84$ and $ZJ_2=0.68$.  They corresponds to four red dots in Fig.\ref{Fig:phase}(a). Linear dotted lines in (b) and  (c) represent the velocity of a NG mode. In (b) a dash-dot line represents 
the 1st excitation band away from the SS/SF boundary, where an arrow indicates 
how the dispersion dips into the origin.  (e) The velocity of Nambu-Goldstone mode against $h$.  
Blue dots are derived directly from the excitation spectra, while red dots 
represent Eq.(\ref{eq:velocity}). (f) The original 
Brillouin zone (${\bf BZ}$; large square) for the phases without $Z_2$ breaking symmetry, and the folded Brillouin zone (${\bf BZ}/2$; blue area) 
for phases with the broken symmetry are indicated. 
The dotted line shows the $k_x=k_y=k$ direction.}
  \label{Fig:excitation}
\end{figure}

If we plug this into the Hamiltonian, and 
neglect the terms with more than two boson operators (which amounts to 
the linear spin wave approximation), the effective Hamiltonian takes 
a form, $H_{\mathrm{eff}}=H_{\mathrm{eff}\pm}+H_{\mathrm{eff}0}$, where $H_{\mathrm{eff}\pm}$ composed of $b^{\dagger}_{\pm}$ and $H_{\mathrm{eff} 0}$ composed of $b^{\dagger}_{00}$.
The form of $H_{\mathrm{eff}\pm}$ is, up to a constant, 
\begin{equation}
H_{\mathrm{eff}\pm} = \frac{1}{2}\sum_{{\bf k}\in {\bf BZ}/2}\futomoji{\psi}^{\dagger}_{{\bf k}}\hat{H}_{\mathrm{eff}\pm}({\bf k})
 \futomoji{\psi}_{{\bf k}},
\label{eq:spinwave}
\end{equation}
where {\small $\futomoji{\psi}_{{\bf k}}=(b^{A}_{{\bf k},+},b^{A}_{{\bf k},-},b^{B}_{{\bf k},+},b^{B}_{{\bf k},-},
b^{A\dagger}_{{-\bf k},+},b^{A\dagger}_{-{\bf k},-},b^{B\dagger}_{-{\bf k},+},b^{B\dagger}_{-{\bf k},-})^T$}, and $\hat{H}_{\mathrm{eff}\pm}$ is an $8\times8$ matrix, whose components are shown in Appendix.  The folded Brillouin zone ${\bf BZ}/2$ denotes  the 1st Brillouin zone when the symmetry between A and B sublattices is broken, see Fig.\ref{Fig:excitation}(f). 
We can diagonalize the effective Hamiltonian as
\begin{equation}
H_{\mathrm{eff}\pm} = \sum_{{\bf k}\in {\bf BZ}/2,\tau} \epsilon_{\tau}({\bf{k}})\beta^{\dagger}_{{\bf k},\tau}\beta_{{\bf k},\tau},
\label{eq:diag}
\end{equation}
with the band index $\tau=1-4$.  Here we have applied a Bogoliubov transformation $U$: $\futomoji{\psi}'_{{\bf k}}=U\futomoji{\psi}_{{\bf k}}$, 
where $U$ is an $8\times8$ matrix satisfying $U\Sigma U^T=\Sigma$ 
with $\Sigma=\mathrm{diag}(1,1,1,1,-1,-1,-1,-1)$, and $\futomoji{\psi}'_{{\bf k}}=(\beta_{{\bf k},1},\beta_{{\bf k},2},\beta_{{\bf k},3},\beta_{{\bf k},4},\beta^{\dagger}_{{-\bf k},1},\beta^{\dagger}_{-{\bf k},2},\beta^{\dagger}_{-{\bf k},3},\beta^{\dagger}_{-{\bf k},4})^T$\cite{Hubbarddynam1}.  
Note that the above is applicable to SF and MI phases.  There, the excitation spectrum appears to have 4 bands since we treat them in the folded Brillouin zone ${\bf BZ}/2$. When we unfold the Brillouin zone into the full {\bf BZ}, they have 2 bands. Note that the main difference between the results in SS and those in SF within this method is that there is no degenerated modes at the boundary of ${\bf BZ}/2$ in SS. We also note that one can evaluate the expectation value of physical quantities in this approximation as follows. An operator representing a quantity is first transformed with $b$. Then the HP transformation is applied to $b^{\lambda}_0, b^{\lambda \dagger}_0$, and the expansion is made up to the order needed.  
When $b^{\lambda}_0, b^{\lambda\dagger}_0$ appear in the form of $n_{\lambda0}$, this can be dealt with using the constraint directly. Finally, we take the expectation value for the transformed operator.

 In Fig.\ref{Fig:excitation}(a-d), we show the excitation spectrum in the phases MI, SF, SS, CO, respectively, for which 
the positions on the phase diagram 
are indicated in Fig. \ref{Fig:phase} (a). The spectrum is drawn along 
the $(1,1)$ direction ($k_x=k_y$).  Here we only show the excitations composed of $b_{\pm}$, since only these modes have the counterparts in the bose Hubbard model. In MI and SF, there are two excitation bands. On the other hand, in SS and CO, there are four bands because the symmetry between A,B sublattices is broken. There is a gap in MI ($n=1$), and the two (three if $t_0$ is included) modes are degenerate at $h=0$. 
As $h$ increases, the degeneracy is lifted 
due to the Zeeman splitting (Fig.\ref{Fig:excitation}(a)). The dispersion relation is given by 
\begin{equation}
\epsilon({\bf k})=\sqrt{J_0^2+2J_0J_2\gamma({\bf k})}\pm h, \label{eq:mi_dis}
\end{equation}
where $\gamma({\bf k})=\sum_{a=1}^{d}\cos(k_a)$ with $a$ labeling the 
axes.  

 When the gap closes, a quantum phase transition  between SF and MI occurs. In the SF phase, there is one gapless mode (a Nambu-Goldstone (NG) mode), which 
arises from the U(1) symmetry breaking. In particular, the velocity of NG mode 
vanishes at the boundary except for $h=0, ZJ_2=J_0$. As $h$ becomes closer to the boundary of SF and SS, a dip (i.e., softening) appears in the mode around 
${\bf k}=(0,0)$, and we observe that the gap closes 
at the boundary (Fig.\ref{Fig:excitation}(b)).  This mode can be thought as a roton mode, which represents a softening into CO or SS. Note that the roton is located at the zone center rather than a boundary, since the 
hopping parameter is positive in our model.  
If we turn to the excitation spectrum of SS state in Fig.\ref{Fig:excitation}(c), there is one NG mode with a linear dispersion. The velocity of the NG mode at the boundary of SS and CO becomes 0. In the CO phase, there is an energy gap, which is closed at the boundary between CO and SS, and the highest mode is flat 
with $\epsilon({\bf k})=2h$, Fig.\ref{Fig:excitation}(d). In this mode up-spin 
triplet flips into a down-spin triplet. The rest of the excitation spectrum can be obtained by solving an equation cubic  in $x$ for each ${\bf k}$,
 \begin{equation}
 \begin{split}
 0=&[x+h-J_0-Z(J_1+J_2/2)](x+h-J_0)\\&\times[x+h+J_0-Z(J_1+J_2/2)]+2J_0
J_2^2\gamma({\bf k})^2.
 \end{split}
 \end{equation}
 Strictly speaking, the actual excitation corresponds to the absolute value of the solution. From the structure of the equation and straightforward manipulation, we notice that the dispersion does not change against $h$, 
and the energy of one of the bands decreases as $h$ increases. Therefore, there occurs a band crossing at some $h$ in CO. In addition, one can show analytically that a band never has linear dispersion at the point where the gap in CO spectrum is closed.  
 This contrasts with the case of MI ($n=1$), where, at $h=0$ and $ZJ_2=J_0$, the gap is closed but the band has a linear dispersion, see Eq.(\ref{eq:mi_dis}).
 It is numerically confirmed that, at the SF/CO boundary, the gap in the CO phase does not close, nor does the roton mode. (Near the CO/SF boundary, a roton mode appears again.) The velocity of the NG mode approaches zero toward the boundary of SF and MI ($n=2$). In the latter phase, the upper band is flat [$\epsilon({\bf k})=2h-2Z(J_1+J_2/2)$]. The analytic  expression for the other band is 
\begin{equation}
\epsilon({\bf k})=-J_0+h-Z\left(J_1+\frac{J_2}{2}\right)+J_2\gamma({\bf k}).
\end{equation}

To gain further understanding of the excitations, 
we can actually obtain an analytic expression for the
spin wave velocity (with the derivation given in Appendix), 
\begin{equation}
C_s = \left(\frac{J_2}{2\kappa} |M_{xy,A}M_{xy,B}|\right)^{1/2},
\label{eq:velocity}
\end{equation}
where $\kappa=\partial m_z/\partial h$ denotes the spin susceptibility, and $M_{xy,A}, M_{xy,B}$ the magnetization of each sublattice. Note that $M_{xy}$ and $m_z$ here do not include spin wave corrections, 
which is not negligible in two dimensions. 
This microscopic expression agrees with the relation derived from phenomenological discussions (hydrostatically or with an effective Lagrangian)
for spin systems\cite{spinwavevelocity1} 
and for  bose systems\cite{bosevelocity1,bosevelocity2}.  
It is also similar to the result obtained by the Gutzwiller approximation 
for SF\cite{Hubbarddynam2}.  We can also 
show that the above expression holds in SS when the Gutzwiller approximation is used for the extended bose Hubbard model.  
The velocity plotted against $h$ from Eq.(\ref{eq:velocity}) 
 is displayed in Fig.\ref{Fig:excitation}(e) along with the numerical result, 
and we can see the two sets of results almost exactly coincide with each other.  
Then, a jump in the velocity at the SF/SS boundary can be attributed to a jump in the spin susceptibility.  
The velocity vanishes at MI/SF and SS/CO boundaries, since $M_{xy}$ becomes 0. 
 \begin{figure}[h]
   \includegraphics[width=90mm]{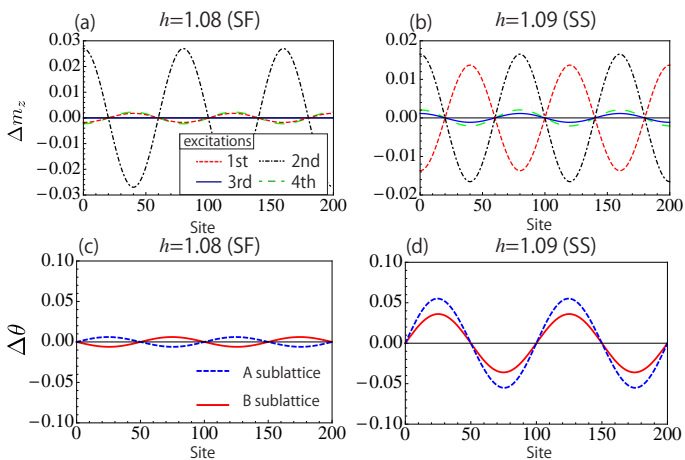}
  \caption{ Spatial variation of physical quantities for excitation modes for small $k$ in {\bf BZ}/2. We have chosen parameter $ZJ_1=0.84$ and $ZJ_2=0.68$ . (a)(b) show the modulation of the density ($\Delta m_z$) in A sublattice at  $h=1.08$ (SF) and h=$1.09$ (SS), respectively. 
  The red line represents the first excited mode(a NG mode), 
the black line  the second excitation (roton) mode, the blue line the third excitation (massive) mode, and the green line the fourth mode.
(c)(d) show the spatial variation of the phase of $M_{xy,A}$ and $-M_{xy,B}$ for  the second excitation mode for $h=1.08$(c) or $h=1.09$(d).}
     \label{Fig:prop1}
  \end{figure}
 
Next, we discuss the properties of collective excitation modes. 
We focus on all four modes for small ${\bf k}$ along the (1,1) direction in the folded Brillouin zone {\bf BZ}/2.  
(For SF, a NG mode, a roton mode, a massive mode and the rest are investigated.) To reveal the character of each excitation ($\beta^{\dagger}_{{\bf k},\tau}$), we construct a coherent state ($|\chi_{{\bf k},\tau}\rangle$) to calculate the spatial variation of $m_{z,i}$ (=$\langle S^z_{i,1}+S^z_{i,2} \rangle$) and $M_{xy,i}$.  
In the boson language the former corresponds to spatial 
density modulations, while the latter to modulations of the order parameter 
($\langle a^{\dagger}\rangle$). 
The coherent state is expressed as
\begin{equation}
|\chi_{{\bf k},\tau}\rangle = \exp(-|\chi_{{\bf k},\tau}|^2/2)\exp(\chi_{{\bf k},\tau}\beta^{\dagger}_{{\bf k},\tau})|0\rangle.
\end{equation}
Here we choose $\chi_{{\bf k},\tau}$ to be small, 
which amounts to assuming that there are not 
too many spin waves.   In SF, it turns out that, except for the 3rd excitation mode (a massive mode in Fig.\ref{Fig:excitation}(b) near $k \approx \pi$), excitation modes are accompanied by both a modulation of the order parameter (i.e., superfluid density) and the density. On the other hand, the massive mode does not exhibit modulation in the density but a local 
imbalance between condensate and noncodensate amplitudes, as 
seen in Fig.\ref{Fig:prop1}(a). This agrees with the result for cold 
atoms \cite{Hubbarddynam1}.  In the SS phase, by contrast, such 
a mode disappears, as seen in Fig.\ref{Fig:prop1}(b). All the four modes in SS 
are accompanied by modulations of density and order parameter.
Figures \ref{Fig:prop1}(c),(d) display the spatial variation of the phase (Arg$(M_{xy})$) of $M_{xy,A},\;-M_{xy,B}$ ($\equiv\Delta \theta$) 
 for the second excitation mode (the roton mode) at $h=1.08$ (SF) and $h=1.09$ (SS), respectively. 
 Note that the minus sign in $-M_{xy,B}$ again comes from the antiferromagneic coupling $J_2$.
 We can see that the relation of the phase modulation between A and B sublattice is different between SF and SS.  
In SF, the roton mode may be thought of as a Leggett mode if we 
regard one cell as composed of two neighboring  sites (one belongs to A sublattice, and the other to B), since 
the phase of the order parameter out phase between A and B sublattices.
On the other hand, this interpretation cannot be applied to SS where
the phase modulation is in phase between them.  
We have to note that this property of the second excitation in the SS phase changes for large enough $V$ and away form the SS/SF boundary. Then the 
phase modulation becomes out of phase between A and B sublattices (as in SF). As for the NG mode, the spatial modulation of the phase of the order parameter (not shown) is in phase between A and B sublattices in both of the states.

\section{Spin-Spin correlation}
Analysis of spin-spin correlations is important from the experimental viewpoint, since 
inelastic neutron scattering can detect it. Here we focus on two kinds of spin-spin correlations that have counterparts in bose systems. The first one is 
\begin{equation}
\begin{split}
C^{z}({\bf k},\omega)&\equiv \int^{\infty}_{-\infty}d\tau e^{i\omega \tau}\langle S^{z}_{\bf k}(\tau)S^{z}_{-\bf k}(0)\rangle \\
&=\sum_{n}|\langle n|S^{z}_{-{\bf k}}|0\rangle|^2 \delta(\omega-\epsilon_n),
\label{eq:cz}
\end{split}
\end{equation}
where $S^z_{i}\equiv S^z_{1,i}+S^z_{2,i}=t^{\dagger}_{+}t_{+}-t^{\dagger}_{-}t_{-}$, $|0\rangle$ denotes the ground state, $|n\rangle$ an excited state 
with an energy $\epsilon_n$. The correlation function corresponds to the dynamical structure factor in cold atom systems, which can be
detected with Bragg spectroscopy. 
The second correlation function is 
\begin{equation}
\begin{split}
C^{+-}({\bf k},\omega)&\equiv \int^{\infty}_{-\infty}d\tau e^{i\omega \tau}\langle \hat{M}_{xy, -{\bf k}}(\tau)\hat{M}^{\dagger}_{xy, -{\bf k}}(0)\rangle \\
&=\sum_{n}|\langle n|\hat{M}^{\dagger}_{xy, -{\bf k}}|0\rangle|^2 \delta(\omega-\epsilon_n).
\label{eq:cpm}
\end{split}
\end{equation}
Here $\hat{M}_{xy, i}\equiv (S^{+}_{1,i}-S^{+}_{2,i})/\sqrt{2}=t^{\dagger}_{+i}s_i+s^{\dagger}_it_{i-}$, which is an operator form of ${M}_{xy, i}$ introduced 
before. This correlation function corresponds to the lesser Green's function $G^{<}_{-{\bf k}}(\omega)\equiv-i\int^{\infty}_{-\infty}d\tau e^{i\omega \tau}\langle a^{\dagger}_{-\bf k}(\tau)a_{-\bf k}(0)\rangle$, where $a$ is a bosonic annihilation operator.  
In the spin wave theory, there are $n$-spin wave states defined as $(\beta^{\dagger})^n|\mathrm{vac}\rangle$.  
In the following, we focus on the intensity of the single spin wave peak 
by taking $|{\bf k},\tau\rangle = \beta^{\dagger}_{{\bf k},\tau}|\mathrm{vac}\rangle$ as $|n\rangle$, where $\beta^{\dagger}_{{\bf k},\tau}$ is defined 
in eqn.(\ref{eq:diag}).  
Within the spin wave theory, one can evaluate the coefficient of the delta-function ($\delta(\omega-\epsilon_n)$ in Eq.(\ref{eq:cz}),(\ref{eq:cpm})) as 
\begin{equation}
\begin{split}
&\langle \tau,-{\bf k}_0|S^{z}_{-{\bf k}}|0\rangle\\
&=\sqrt{\frac{N}{2}}\sum_{\lambda}\{ v_{\lambda}u_{\lambda}(f_{\lambda}^2-g_{\lambda}^2)e^{i{\bf G\cdot l_{\lambda}}}\\
&\times[N_{({\lambda},+)\tau}({\bf k}_0)+P_{({\lambda},+)\tau}({\bf k}_0)]\\
&-2v_{\lambda}g_{\lambda}f_{\lambda}e^{i{\bf G}\cdot{\bf l}_{\lambda}}[N_{({\lambda},-)\tau}({\bf k}_0)+P_{({\lambda},-)\tau}({\bf k}_0)] \}
\end{split}
\end{equation}
and
\begin{equation}
\begin{split}
&\langle \tau,-{\bf k}_0|\hat{M}_{xy, -{\bf k}}^{\dagger}|0\rangle=\\
&\sqrt{\frac{N}{2}} \sum_{\lambda}e^{i{\bf G}\cdot{\bf l}_{\lambda}}
[(u_{\lambda}^2g_{\lambda} -v_{\lambda}^2f_{\lambda})N_{(\lambda,+),\tau}
({\bf k}_0) \\&+(u_{\lambda}^2f_{\lambda} -v_{\lambda}^2g_{\lambda})P_{(\lambda,+),\tau}({\bf k}_0)\\
&+u_{\lambda}f_{\lambda}N_{(\lambda,-),\tau}({\bf k}_0)-u_{\lambda}g_{\lambda} P_{(\lambda,-),\tau}({\bf k}_0)],
\end{split}\label{eq:pm_detail}
\end{equation}
where ${\bf l}_{\lambda}$ is an arbitrary site in the sublattice $\lambda$, 
${\bf G}$ is $(\pm \pi,\pm \pi)$ or $(0,0)$ chosen so that ${\bf k}_0$ be in {\bf BZ}/2, and ${\bf k}={\bf k}_0+{\bf G}$. 
$P({\bf k}), N({\bf k})$ are the 
elements of the matrix used for the 
Bogoliubov transformation (Eq.(\ref{eq:diag})),
\begin{equation}
U({\bf k})=\begin{pmatrix} N({\bf k}) & P({\bf k}) \\ P({\bf k}) & N({\bf k})
\end{pmatrix}
,
\end{equation}
where the elements of $4\times4$ $N, P$ are denoted as 
$N_{(\lambda,-),\tau}, P_{(\lambda,\theta),\tau}$.
The intensity of a single spin wave mode is expressed as $|\langle \tau,
-{\bf k}_0|S^{z}_{-{\bf k}}|0\rangle|^2$ for $C^{z}({\bf k},\omega)$ and 
$|\langle \tau,-{\bf k}_0|\hat{M}^{\dagger}_{xy, -{\bf k}}|0\rangle|^2$ for 
$C^{+-}({\bf k},\omega)$.

\begin{figure}[h]
  \begin{center}
   \includegraphics[width=90mm]{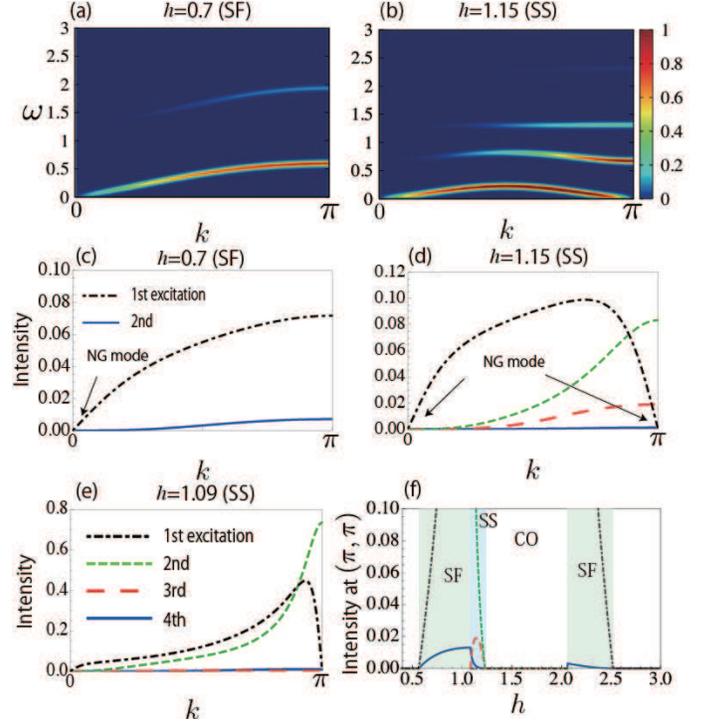}
  \end{center}
  \caption{{\small (a)(b) Color-coded intensity of $C^z({\bf k},\omega)$  
against $k_x=k_y=k$ 
for a SF phase with $h=0.7$ (a) and for a SS phase with $h=1.15$ (b), 
with $ZJ_1=0.84$ and $ZJ_2=0.68$. A gaussian of width $\delta = 0.03J_0$ is used to smooth the $\delta$-functions in $C^z({\bf k},\omega)$. (c-f) The intensity of peaks of $C^z({\bf k},\omega)$ against $k_x=k_y=k$ for single spin wave states for 
$h$=0.7 (c),  $h$=1.15 (d), and $h$=1.09 (e), with $ZJ_1=0.84$ and $ZJ_2=0.68$. For the homogeneous states, the 1st (2nd) band is represented by black (blue) lines. For the state with broken $Z_2$ symmetry, blue, red, green, and black lines represent the 1st, 2nd, 3rd and 4th excitations, respectively. (f) The intensity against $h$ at $(k_x,k_y)=(\pi,\pi)$.  
Note the difference in the scale of the vertical axis 
between panels (c)-(f).
} }
  \label{Fig:spin-spin}
\end{figure}
Figure \ref{Fig:spin-spin} shows the intensity plot of $C^z({\bf k},\omega)$ ((a-b)) and the intensity of each peak (i.e., $|\langle \tau,-{\bf k_0}|S^{z}_{-{\bf k}}|0\rangle|^2/N$, see (c-f)) along $(k_x,k_y)=(k,k)$. First, we have to note for CO and MI that there is no contribution from the one spin wave excitation modes, i.e., $\langle \tau,-{\bf k_0}|S^{z}_{-{\bf k}}|0\rangle=0$. 
For SF, two excitation bands contribute to the spin-spin correlation as seen in Fig.\ref{Fig:spin-spin}(a)(c), since there is no band folding. Around 
${\bf k}=(0,0)$, the excitations consist of a Nambu-Goldstone (NG) mode (whose intensity grows linearly with $k$), 
and a massive mode (whose intensity grows as $k^4$), and we also 
find a roton mode around ($\pi,\pi$). Such behaviors match those of the dynamical structure factor in the boson systems \cite{Hubbarddynam1,Hubbarddynam2}. As can be seen in Fig.\ref{Fig:spin-spin}(f), which shows the intensity of peak against $h$ at the zone boundary ${\bf Q}=(\pi,\pi)$, the intensity rapidly increases toward the SF/SS or SF/CO boundaries. 
This can be regarded as a hallmark for the phase transition for SS or CO.  

If we turn to the SS state in Fig.\ref{Fig:spin-spin}(b),(d),(e), there are four single particle excitations that contribute to the spin-spin correlation.  
Since the $Z_2$ symmetry is broken in SS, the selection rule for the 
matrix elements appearing in eqns(\ref{eq:cz},\ref{eq:cpm}) is the 
same for ($k_x, k_y$) as that for ($k_x+\pi, k_y+\pi$), so that the NG mode 
in  Fig.\ref{Fig:spin-spin}(b) appears 
both around $(0,0)$ and $(\pi,\pi)$.  
It turns out that the intensity of the 4th band in the SS phase 
rapidly decreases 
toward the boundary to CO. The intensity of this band  is  much weaker than 
those for the other bands in the SS region. At long wave lengths ($k\sim 0$), the intensity also starts from 0 in most part of the SS state. Only the NG mode has an 
intensity increasing 
linearly with $k$, while the other three modes increase like $k^4$. Just after the transition to the SS state in Fig.\ref{Fig:spin-spin}(e), the dominant 
excitations are the lowest two modes. Away from the phase boundary in  Fig.\ref{Fig:spin-spin}(d), the intensities of the lowest three modes become comparable 
with each other. The fact that the massive mode (the 3rd excitation mode around (0,0)$\in${\bf BZ}/2)) has a significant intensity for the $C^z$ correlation function reflects the property of the mode that it becomes coupled with the density modulation in SS. We also find that the intensity of the NG mode vanishes at the zone boundary in SS. This reflects the fact that zero-energy excitation is only coupled with the phase oscillation of the order parameter. Another characteristic property is that the intensity is stronger around ($\pi,\pi$) than around ($0,0$) when we compare ($k_x, k_y$)$\in${\bf BZ}/2 and ($k_x+\pi, k_y+\pi$)$\not\in${\bf BZ}/2. Therefore the  bands can be observed more clearly around ($\pi,\pi$) 
as seen in Fig.\ref{Fig:spin-spin}(b). 
\begin{figure}[h]
  \begin{center}
   \includegraphics[width=90mm]{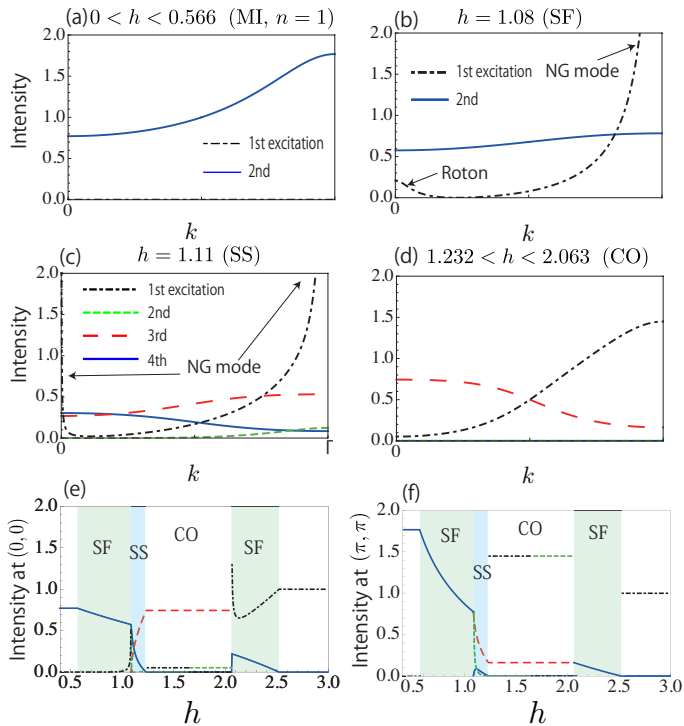}
  \end{center}
  \caption{{\small 
The intensity of peaks  against $k_x=k_y=k$ 
for single spin wave states in the $C^{+-}({\bf k},\omega)$ correlation 
with $ZJ_1=0.84$ and $ZJ_2=0.68$. (a) is for 
$0<h<0.566$ (MI, $n=1$), (b) $h=1.08$ (SF), (c) $h=1.11$ (SS), 
and (d) $1.232<h<2.063$ (CO). For the homogeneous states, the 1st and the 2nd bands are 
represented by black and blue lines, respectively. For the state with $Z_2$ symmetry broken, the blue, red, green, and black lines mean the 1st, 2nd, 3rd and 4th excitations, respectively. (e) shows the intensity against $h$ at $(k_x,k_y)=(0, 0)$, and (f) for at $(k_x,k_y)=(\pi,\pi)$ } }
  \label{Fig:spin-spin_pm}
\end{figure}

Finally we discuss the behavior of the correlation 
$C^{+-}$. 
Figure \ref{Fig:spin-spin_pm} shows the intensity ($|\langle \tau,-{\bf k_0}|\hat{M}^{\dagger}_{xy, -{\bf k}}|0\rangle|^2/N$) for $C^{+-}$ along $(k_x,k_y)=(k,k)$. As seen in Eq.(\ref{eq:pm_detail}), single spin wave states contribute 
to the matrix element in the insulating phases (MI, CO). It turns out that in MI($n=1$) in Fig.\ref{Fig:spin-spin_pm}(a), 
only the second excitation band has a significant intensity, 
which does not depend on $h$ within the present approximation. In the CO state 
in Fig.\ref{Fig:spin-spin_pm}(d), we can see two bands in the spectrum, i.e., the 1st band (the 2nd at higher $h$) and the 3rd band (Fig.\ref{Fig:spin-spin_pm}(d)). 
Namely, as $h$ increases in CO, the two bands change their orders, hence 
the nature of the 1st and 2nd bands changes. This can be seen in  Fig.\ref{Fig:spin-spin_pm}(e)(f), where the intensity of the 1st band is finite at lower $h$ in CO, while the intensity of the 2nd band is finite at higher $h$. In MI($n=2$), only the first band is observed, whose intensity is $|\langle \tau,-{\bf k}_0|\hat{M}_{xy, -{\bf k}}^{\dagger}|0\rangle|^2/N=1$ throughout (not shown). In the SF phase, both bands have contributions to $C^{+-}$. An interesting observation is that the intensity to the NG mode diverges like $1/|(k_x-\pi,k_y-\pi)|$ toward $(\pi,\pi)$. 
Therefore, the NG mode is expected to be observed clearly. As for the roton mode, its intensity increases near the boundary of SF/SS or SF/CO, see Fig.\ref{Fig:spin-spin_pm}(b)(e). This increase is a precursor to the breakdown of $Z_2$ symmetry. In the SS phase in Fig.\ref{Fig:spin-spin_pm}(c), the NG mode becomes 
intense toward ${\bf k}=(0,0), (\pi,\pi)$. In this phase, the intensity of the 1st band diverges like $1/|(k_x,k_y)|$ at ${\bf k}=(0,0)$ and like 
$1/|(k_x-\pi,k_y-\pi)|$ at ${\bf k}=(\pi,\pi)$. 
On the other hand, it turns out that the intensities of the 2nd to 4th bands in the SS phase strongly depend on ($J_1,J_2,h$). 
Given the correspondence between the bose Hubbard model and the spin system, these characteristic behaviors in SS revealed here 
can be expected in cold atoms in optical lattice as well.

\section{ Conclusion}
In this paper we have studied the frustrated spin dimer model with the bond operator and the generalized spin wave theory.
 First we have obtained the phase diagram, which can be directly compared with that of the extended bose Hubbard model. We have revealed how the supersolid state emerges in this model, and found that there is a phase (``SF") which involves $|1,0\rangle$ state and hence has no counterpart in the bose Hubbard model. In addition, we point out that enlarged SS region for large V ($\geq ZU$) in the bose 
Hubbard model does not occur in the spin model, because ``SF" phase takes over 
instead.  

Second, we have obtained the excitation modes, especially the excitation spectra around the SS regions. We have found that the dip corresponding to a roton mode completely softens at the boundary of SS, and that a Nambu-Goldstone mode has a linear dispersion even in the SS state.  
We have also microscopically derived within the generalized linear spin wave theory an analytical relation between the velocity of the NG mode, order parameter and spin susceptibility, which agrees with the relation derived from hydrostatic treatments. As for the properties of the excitation modes, it turns out that, in the SF state, the massive mode does not couple with density modulations, while in the SS state there is a coupling. 

Third, we have calculated the spin-spin correlations $C^z$ and $C^{+-}$. $C^z$ corresponds  in the bose system to the dynamical structure factor, while $C^{+-}$  to the lesser Green's function.  In the SF state, the behavior of $C^z$ is similar to its counterpart for cold atoms. 
The resonance of the intensity of the roton mode is thought to be a precursor 
to the breaking of $Z_2$ symmetry (phase transition to CO or SS). In the SS state, the intensity of the NG (sound) mode peak vanishes at the boundary of the 1st Brillouin zone, while the 3rd excitation band becomes significant, which 
reflects the property of the difference in the excitation between SS and SF 
mentioned above. 
As for $C^{+-}$, the spin wave theory predicts that some of the bands have significant 
intensities in the insulating phases (MI,CO). On the other hand, the NG mode is expected to be clearly observed in the spin-spin correlation, while the intensity of the roton mode can be regard as an evidence for the phase transition to CO or SS. The properties of SS revealed here for the spin model should be applicable to the SS phase in bose systems.  
Hence the results should be important probes in searching SS in a wide range of 
systems.

\section{Acknowledgements}
We wish to thank D. Yamamoto and T. Morimoto for helpful discussions.
\section{ Appendix A}

Let us here display an $8\times8$ matrix 
$\hat{H}_{\mathrm{eff}}({\bf k})$ explicitly, whose form is 
\begin{equation}
\hat{H}_{\mathrm{eff}}({\bf k})=
\begin{bmatrix} 
E^{A}_{+} & E^{A}_{\pm} & C_{+} & C'_{\pm} &0 &0 &D_{+}&D_{\pm} \\
E^{A}_{\pm} & E^{A}_{-} & C_{\pm} & C_{-} &0 &0 &D'_{\pm}&D_{-}\\
C_{+} & C_{\pm} &E^{B}_{+}&E^{B}_{\pm}&D_{+}&D'_{\pm}&0&0\\
 C'_{\pm}&C_{-} &E^{B}_{\pm}&E^{B}_{-}&D_{\pm}&D_{-}&0&0\\
 0&0&D_{+}&D_{\pm} &E^{A}_{+} & E^{A}_{\pm} & C_{+} & C'_{\pm}\\
 0 &0 &D'_{\pm}&D_{-}&E^{A}_{\pm} & E^{A}_{-} & C_{\pm} & C_{-}\\
 D_{+}&D'_{\pm}&0&0&C_{+} & C_{\pm} &E^{B}_{+}&E^{B}_{\pm}\\
 D_{\pm}&D_{-}&0&0& C'_{\pm}&C_{-} &E^{B}_{\pm}&E^{B}_{-}
 \end{bmatrix}.
\end{equation}
Expressions for some of the more complicated elements are 
\begin{equation}
 \begin{split}
  E^{\lambda}_{\pm} = &2hg_{\lambda}u_{\lambda}f_{\lambda} - 2\left(J_1 + \frac{J_2}{2}\right)Zv_{\bar{\lambda}}^2u_{\lambda}f_{\lambda}g_{\lambda}(f_{\bar{\lambda}}^2 - g_{\bar{\lambda}}^2)\\& + 
  \frac{J_2}{2}Zv_{\bar{\lambda}}v_{\lambda}u_{\bar{\lambda}}(f_{\bar{\lambda}}g_{\lambda} - g_{\bar{\lambda}}f_{\lambda}- f_{\lambda}f_{\bar{\lambda}} + g_{\lambda}g_{\bar{\lambda}}),
  \end{split}
  \end{equation}
 \begin{equation}
 \begin{split}
  C_{\pm} = &J_2\gamma({\bf k})[u_Au_B^2(f_Ag_B - f_Bg_A) + v_B^2u_A(g_Ag_B - f_Af_B)] \\
  &-4\left(J_1 + \frac{J_2}{2}\right)\gamma({\bf k})g_Af_Av_Av_Bu_B(f_B^2 - g_B^2),
  \end{split}
  \end{equation}
and
  \begin{equation}
\begin{split}
D_{\pm} = &J_2\gamma({\bf k})[v_A^2u_B(f_Ag_B - g_Af_B) + u_Bu_A^2(f_Af_B - g_Ag_B)] \\
&-4 \left(J_1 + \frac{J_2}{2}\right)\gamma({\bf k})g_Bf_Bv_Bv_Au_A(f_A^2 - g_A^2).
  \end{split}
  \end{equation}
Here $Z$ is the coordination number, $\gamma({\bf k})=\sum_{a=1}^{d}\cos(k_a)$, $\lambda=A,B$, and  $\bar{\lambda}$ denotes the sublattices excluding  $\lambda$.\\

\section{ Appendix B: Derivation of Eq.(\ref{eq:velocity})}
We show how we can derive an analytic expression, Eq.(\ref{eq:velocity}).  
The idea is, if we are only interested in the velocity, we can make use of the equations governing the coefficients for bosons ($t^{\dagger},s^{\dagger}$) in the ground state. The similar idea is used in the context of Gutzwiller approximation for boson models \cite{Hubbarddynam2}.  
The proof consists of three steps. 
In the 1st step, we introduce another way to derive the excitation. In the 2nd step, we show that the resultant excitation spectrum is the same as that in the main text. In the 3rd step, we prove Eq.(\ref{eq:velocity}) within the approach introduced in the 1st step. We start from Eq.(\ref{eq:effective1}), which neglects the existence of $|1,0\rangle $, and uses the language of bose systems ( Eq.(\ref{eq:corresponde})) to characterize parameters in the Hamiltonian. Note that, in the following, we also change $s^{\dagger}\rightarrow t^{\dagger}_{0}$ and $y\rightarrow x_{0}$ to simplify the notation.
\subsection{1st step}
The original assumption is that the form of the ground state is $\prod_{i\in A} (x_{A,1}t_{i+}^{\dagger}+x_{A,0}t_{i0}^{\dagger}+x_{A,-1}t_{i-}^{\dagger})\prod_{i\in B} (x_{B,1}t_{i+}^{\dagger}+x_{B,0}t_{i0}^{\dagger}+x_{B,-1}t_{i-}^{\dagger})|{\rm vac}\rangle$, whose norm is 1. We extend this to assume that the dynamics is confined to this type of states, and that when we consider the dynamics of site $g$ the effect of the surrounding state can be regarded as a mean field 
(which is an idea similar to the Gutzwiller approach for bose Hubbard model \cite{Hubbarddynam2}).  In other words, to consider the dynamics of the state on site $g$ at time $\tau$, we use the local Hamiltonian,
\begin{equation}
{\scriptstyle
\begin{split}
H_{g}(\tau)&=-t\sum_{i_g}
[(t_{g+}^{\dagger}t_{g0}+t_{g0}^{\dagger}t_{g-})\phi_{ i_g}(\tau)+{\rm H.c.}]\\
&+V\sum_{i_g}(t_{g+}^{\dagger}t_{g+}-t_{g-}^{\dagger}t_{g-})\delta n_{i_g}(\tau)\\
&+\frac{U}{2}(t_{g+}^{\dagger}t_{g+}+t_{g-}^{\dagger}t_{g-})-h(t_{g+}^{\dagger}t_{g+}-t_{g-}^{\dagger}t_{g-}),
\end{split}
}
\end{equation}
where $i_g$ stands for the nearest neighbors of $g$, $\phi_i(\tau)=\langle t_{i0}^{\dagger}t_{i+}+t_{i-}^{\dagger}t_{i0}\rangle=x_{i,0}^{*}(\tau)x_{i,1}(\tau)+x_{i,-1}^{*}(\tau)x_{i,0}(\tau)$ and $\delta n_{i}(\tau)=\langle t_{i+}^{\dagger}t_{i+}-t_{i-}^{\dagger}t_{i-}\rangle=x_{i,1}^{*}(\tau)x_{i,1}(\tau)-x_{i,-1}^{*}(\tau)x_{i,-1}(\tau)$. Then the equation of motion for the coefficient is 
\begin{equation}
\begin{split}
i\frac{d x_{g,\theta}(\tau)}{d\tau}&=-t\sum_{ \sigma,\theta'}[\phi_{g+\sigma}\delta_{\theta,\theta'+1}+\phi^*_{g+\sigma}\delta_{\theta,\theta'-1}]x_{g,\theta'}(\tau)\\
&+\left[\frac{U}{2}\theta^2-\theta h+\theta V(\sum_{\sigma}\delta n_{g+\sigma})\right]x_{g,\theta}(\tau), 
\end{split}
\label{eom_1st}
\end{equation}
where $\theta=\pm1,0$.
In the ground state, the local Hamiltonian can be different between A, B sublattices, which we express as $H_{A}, H_{B}$. We can express the variational ground state as $\prod_i(\sum_{\theta}d^{0\lambda}_{i,\theta}t^{\dagger}_{i,\theta})|{\rm vac}\rangle$($i\in\lambda$), where $d^{0\lambda}_{i,\theta}$ is the 
optimized parameter.  Then the state on sublattice $\lambda$ ($\sum_{\theta}d^{0\lambda}_{\theta}t^{\dagger}_{\theta})|{\rm vac}\rangle$) is an eigenstate of $H_{\lambda}$ with an eigenvalue $\omega_{\lambda}$. 
We derive the excitation spectrum by considering the stationary solutions around the ground state. In order to do this, we consider $x^{\lambda}_{g,\theta}(\tau)=[d^{0\lambda}_{g,\theta}+d'^{\lambda}_{g,\theta}(\tau)]\exp(-i\omega_{\lambda}\tau)$, where we have defined $d'^{\lambda}_{g,\theta}(\tau)$ which 
is assumed to be small. The equation of motion  is linearized with respect to $d'^{\lambda}_{g,\theta}(\tau)$, where we consider the stationary solutions with a form
\begin{equation}
d^{'\lambda}_{{\bf l},\theta}(\tau)=u^{\lambda}_{\bf k \theta}\exp[i({\bf k} \cdot {\bf l}-\omega_{\bf k}\tau)]+\nu^{\lambda*}_{{\bf k} \theta}\exp[-i({\bf k} \cdot {\bf l}-\omega_{\bf k}\tau)].
\end{equation}
The resultant equation is 
\begin{equation}
\omega_{{\bf k}}\begin{bmatrix} 
\futomoji{u}^{A}\\
\futomoji{u}^{B}\\
\futomoji{\nu}^{A}\\
\futomoji{\nu}^{B}
 \end{bmatrix}=\begin{bmatrix}
 W_{A,A}&W_{A,B}&0&V_{A,B}\\
W_{B,A}&W_{B,B}&V_{B,A}&0\\
0&-V_{A,B}& -W_{A,A}&-W_{A,B}\\
 -V_{B,A}&0&-W_{B,A}&-W_{B,B}
 \end{bmatrix}\begin{bmatrix} 
\futomoji{u}^{A}\\
\futomoji{u}^{B}\\
\futomoji{\nu}^{A}\\
\futomoji{\nu}^{B}
 \end{bmatrix}.\label{eq:gutzexcitation1}
\end{equation}
Here $\mathrm{W_{\lambda,\lambda'},V_{\lambda,\lambda'}}$ are $3\times3$ matrices with $\futomoji{u}^{\lambda}=(u_{{\bf k},1}^{\lambda},u_{{\bf k},0}^{\lambda},u_{{\bf k},-1}^{\lambda})^T$ and $\futomoji{\nu}^{\lambda}=(\nu_{{\bf k},1}^{\lambda}, \nu_{{\bf k},0}^{\lambda}, \nu_{{\bf k},-1}^{\lambda})^T$.
In the following let us denote the $12\times12$ matrix as $\Upsilon({\bf k})$. The elements of this matrix  are 
\begin{equation}
\begin{split}
&W_{\lambda,\lambda,\theta,\theta'} = \left(\frac{U}{2}\theta^2-h\theta-\omega_{\lambda}+\theta V Z \delta n^{0}_{\bar{\lambda}}\right)\delta_{\theta,\theta'}\\
&\;\;\;\;\;\;\;\;\;\;\;\;\;\;\;\;\;\;\;\;\;\;\;\;-Zt\phi^{0}_{\bar{\lambda}}(\delta_{\theta,\theta'+1}+\delta_{\theta,\theta'-1}),\\
&W_{\lambda,\bar{\lambda},\theta,\theta'}({\bf k}) = -t\gamma({\bf k})(d^{0\lambda}_{\theta-1}d^{0\bar{\lambda}}_{\theta'-1}+d^{0\lambda}_{\theta+1}d^{0\bar{\lambda}}_{\theta'+1})\\
&\;\;\;\;\;\;\;\;\;\;\;\;\;\;\;\;\;\;\;\;\;\;\;\;+V\theta \theta'\gamma({\bf k})d^{0\lambda}_{\theta}d^{0\bar{\lambda}}_{\theta'},\\
&V_{\lambda,\bar{\lambda},\theta,\theta'}({\bf k}) = -t\gamma({\bf k})(d^{0\lambda}_{\theta-1}d^{0\bar{\lambda}}_{\theta'+1}+d^{0\lambda}_{\theta+1}d^{0\bar{\lambda}}_{\theta'-1})\\
&\;\;\;\;\;\;\;\;\;\;\;\;\;\;\;\;\;\;\;\;\;\;\;\;+V\theta \theta'\gamma({\bf k})d^{0\lambda}_{\theta}d^{0\bar{\lambda}}_{\theta'},
\end{split}
\end{equation}
where $\delta n^{0}_{\lambda}$ and $\phi^{0}_{\lambda}$ denotes, respectively, 
$\delta n$ and $\phi$ on sublattice $\lambda$ in the ground state.
Note that, to derive Eq.(\ref{eom_1st}) and Eq.(\ref{eq:gutzexcitation1}), $\sum_{\theta}|x_{\theta}|^2=1$ is assumed. Therefore, the solutions of Eq.(\ref{eq:gutzexcitation1}) which satisfy  $\sum_{\theta}|x_{\theta}|^2=1$ are physical. This is why the $12\times12$ matrix in Eq.(\ref{eq:gutzexcitation1}) and the $8\times8$ matrix, $\hat{H}_{\mathrm{eff}}({\bf k})$, give the same results.

\subsection{2nd step}
The idea for deriving above equation of motion does not depend on the choice of the basis employed. 
Let us define another local Hamiltonian for site $g$
\begin{equation}
\begin{split}
H'_{g}&=-t\sum_{i_g}
[(t_{g+}^{\dagger}t_{g0}+t_{g0}^{\dagger}t_{g-})(t_{i_g0}^{\dagger}t_{i_g+}+t_{i_g-}^{\dagger}t_{i_g0})+{\rm H.c.}]\\
&+V\sum_{i_g}(t_{g+}^{\dagger}t_{g+}-t_{g-}^{\dagger}t_{g-})(t_{i_g+}^{\dagger}t_{i_g+}-t_{i_g-}^{\dagger}t_{i_g-})\\
&+\frac{U}{2}(t_{g+}^{\dagger}t_{g+}+t_{g-}^{\dagger}t_{g-})-h(t_{g+}^{\dagger}t_{g+}-t_{g-}^{\dagger}t_{g-}),
\end{split}
\end{equation}
where we have picked up the part of the Hamiltonian Eq.(\ref{eq:effective1}) 
that involves site $g$. 
Then let $\{b'^{\dagger}_{i,\theta}\}$ be a set of bosonic creation operators 
transformed from $\{t^{\dagger}_{i,\theta}\}$, i.e., $b'^{\dagger}_{i,\theta}=\sum_{\theta'}U^{(i)}_{\theta,\theta'}t^{\dagger}_{i,\theta'}$ with $U^{(i)}$ an arbitrary site-dependent unitary matrix, and express a state as $|\psi(\tau)\rangle=\prod_i(\sum_{\theta}\chi_{i,\theta}(\tau)b'^{\dagger}_{i,\theta})|{\rm vac}\rangle$, where $\chi_{i,\theta}$ is a complex coefficient. As far as  $|\psi(\tau)\rangle$ is normalized the equation of motion introduced in the 1st step is expressed as 
\begin{equation}
\begin{split}
i\frac{d}{dt\tau}\chi_{g,\theta}(\tau)&=\langle {\rm vac} |b'_{g,\theta} \prod_{i\neq g}(\sum_{\theta'}\chi^*_{i,\theta'}(t)b'_{i,\theta'}) H'_g\\
&\times\prod_i(\sum_{\theta}\chi_{i,\theta'}(t)b'^{\dagger}_{i,\theta'})|{\rm vac}\rangle, \label{eq:arb}
\end{split}
\end{equation}
 with $b'^{\dagger}=t^{\dagger}$. However the evolution of the state with Eq.(\ref{eq:arb}) does not depend on the choice of $\{b'^{\dagger}_{i,\theta}\}$. As is explained in the following, it turns out that, if we take $b'^{\dagger}=b^{\dagger}$, the equation for the stationary solution around the ground state leads to the same form as the equation for the excitation spectrum within the spin wave theory, see Eq.(\ref{eq:spinwave}). Therefore, the excitation spectrum derived from the method in the 1st step is same as that from the spin wave theory in the main part of this article.
Let us take $b'^{\dagger}$ equal to $b^{\dagger}$ defined in Eq.(\ref{eq:unitrans}). In this case $\prod_i b_{0,i}^{\dagger}|{\rm vac}\rangle$ is the variational ground state, $|\mathrm{GS}\rangle$. 
 Let 
\begin{equation}
\begin{split}
\chi_{i,\theta}(t)&=C_{i,\theta}(t)\exp(-i\omega_{\lambda}t),\\
& C_{i,\theta}(t)=\begin{cases}
                                       1+c'_{i,0}& \theta=0\\
                                       c'_{i,\theta} &\theta \neq 0,
                                       \end{cases}
\end{split}
\end{equation}
where $i\in\lambda$. Linearizing the equation, we obtain
\begin{equation}
\begin{split}
i\frac{d}{d\tau}c'_{g,\theta}(\tau)
&=\sum_{\theta'\neq0} \alpha^{\lambda}_{\theta,\theta'} c'_{g,\theta'}(\tau)\\
&+\sum_{i\neq g,\theta'\neq0}[\beta^{\lambda}_{\theta,\theta'} c'_{i,\theta'}(\tau)+\gamma^{\lambda}_{\theta,\theta'} c'^*_{i,\theta'}(\tau)],
\end{split}\label{eq:gutzeom2}
\end{equation}
where 
\begin{equation}
\begin{split}
&\alpha^{\lambda}_{\theta,\theta'}=\langle {\rm GS}|b_{g,0}^{\dagger} b_{g,\theta}
(H'_g-\omega_{\lambda})b_{g,\theta'}^{\dagger}b_{g,0}|{\rm GS}\rangle,\\
&\beta^{\lambda}_{\theta,\theta'}=\langle {\rm GS}| b^{\dagger}_{g,0} b_{g,\theta} (H'_g-\omega_{\lambda}) b_{i,\theta'}^{\dagger}b_{i,0}|{\rm GS}\rangle,\\
&\gamma^{\lambda}_{\theta,\theta'}=\langle {\rm GS} | b^{\dagger}_{g,0} b_{g,\theta} b^{\dagger}_{i,0} b_{i,\theta'}(H'_g-\omega_{\lambda})|{\rm GS}\rangle,
\end{split}
\end{equation}
with $g\in \lambda$. Strictly speaking, the form of this equation is a bit different from the e.o.m obtained by linearizing Eq.(\ref{eom_1st}), but it turns out that the equation from Eq.(\ref{eom_1st}) and Eq.(\ref{eq:gutzeom2}) give the same solutions
as far as both of the solutions satisfy the normalization condition $\sum_{\theta}|x_{\theta}|^2=\sum_{\theta}|\chi_{\theta}|^2=1$, i.e., $\futomoji{d^0}\cdot\futomoji{d'}$ is purely imaginary, where $\futomoji{d^0}, \futomoji{d'}$ are defined below eqn(\ref{eom_1st}). 

 When we look at Eq.(\ref{eq:gutzeom2}), we notice that the evolution of the coefficients does not depend on the $\theta=0$ component.  Therefore, when we consider the stationary solution around the ground state, we only have to deal with an eigenvalue problem for the $\theta=\pm$ components, and then we can decide the evolution of the $\theta=0$ component. As in the 1st step, we try to find the solution with the form
\begin{equation}
c^{'\lambda}_{i,\theta} = u'^{\lambda}_{\theta}\exp(i{\bf k}\cdot{\bf r}_i-i\omega_{{\bf k}}t)+\nu'^{\lambda*}_{\theta}\exp(-i{\bf k}\cdot{\bf r}_i+i\omega_{{\bf k}}t).
\end{equation}
Then the eigenvalue problem for the $\theta=\pm$ components reads 
\begin{equation}
\omega_{{\bf k}}\begin{bmatrix} 
\futomoji {u'}^{A}\\
\futomoji{u'}^{B}\\
\futomoji{\nu'}^{A}\\
\futomoji{\nu'}^{B}
 \end{bmatrix}=\begin{bmatrix}
 W'_{A,A}&W'_{A,B}&0&V'_{A,B}\\
W'_{B,A}&W'_{B,B}&V'_{B,A}&0\\
0&-V'_{A,B}& -W'_{A,A}&-W'_{A,B}\\
 -V'_{B,A}&0&-W'_{B,A}&-W'_{B,B}
 \end{bmatrix}\begin{bmatrix} 
\futomoji{u'}^{A}\\
\futomoji{u'}^{B}\\
\futomoji{\nu'}^{A}\\
\futomoji{\nu'}^{B}
 \end{bmatrix},\label{eq:gutzexcitation2}
\end{equation}
where $W'$ and $V'$ are $2\times 2$, $\futomoji{u'}^{\lambda}=(u'^{\lambda}_{{\bf k},1},u'^{\lambda}_{{\bf k},-1})^T, \futomoji{\nu'}^{\lambda}=(\nu'^{\lambda}_{{\bf k},1},  \nu'^{\lambda}_{{\bf k},-1})^T$ and 
\begin{equation}
\begin{split}
&W_{\lambda,\lambda,\theta,\theta'}=\alpha^{\lambda}_{\theta,\theta'},\\
&W_{\lambda,\bar{\lambda},\theta,\theta'}({\bf k})=\gamma({\bf k})\beta^{\lambda}_{\theta,\theta'},\\
&V_{\lambda,\bar{\lambda},\theta,\theta'}({\bf k})=\gamma({\bf k})\gamma^{\lambda}_{\theta,\theta'}.
\end{split}
\end{equation}
By directly evaluating $\alpha,\beta$ and $\gamma$, one can see that the matrix in Eq.(\ref{eq:gutzexcitation2}) is nothing but $\sum\hat{H}_{\mathrm{eff}}({\bf k})$. Therefore the method introduced in the 1st step gives the same excitation spectrum as in the spin wave theory introduced in the main part of this paper.\\

\subsection{3rd step}
We can then move on to show the relationship between the velocity of the NG mode, the order parameters and the spin susceptibility in the way introduced in the 1st step.
First, one can derive an important identity,
\begin{equation}
\begin{split}
\Bigl(\frac{U}{2}\theta^2&-h\theta-\omega_{\lambda}+\theta V Z \delta n^{0\bar{\lambda}}\Bigl) d^{0\lambda}_{\theta}\\
&-Zt\phi^{0\bar{\lambda}}\sum_{\theta'}(\delta_{\theta,\theta'+1}+\delta_{\theta,\theta'-1})d^{0\lambda}_{\theta'}=0,
\end{split}
\end{equation}
from the fact that $x^{\lambda}_{\theta}(t)=d^{0\lambda}_{\theta}\exp(-i\omega_{\lambda})$ is a stationary solution of Eq.(\ref{eom_1st}). Then the first derivative of this equation with respect to $h$ is 
\begin{equation}
\begin{split}
\left(\theta+\frac{\partial\omega_{\lambda}}{\partial h}\right)d^{0\lambda}_{\theta}&=\sum_{\theta'}W_{\lambda,\lambda,\theta,\theta'}\frac{\partial d^{0\lambda}_{\theta'}}{\partial h}\\
&+\sum_{\theta'}[W_{\lambda,\bar{\lambda},\theta,\theta'}({\bf 0})+V_{\lambda,\bar{\lambda},\theta,\theta'}({\bf 0})]\frac{\partial d^{0\bar{\lambda}}_{\theta'}}{\partial h}.\label{eq:key}
\end{split}
\end{equation}
This is the key equation. We can readily find three solutions of Eq.(\ref{eq:gutzexcitation1}) at ${\bf k}={\bf 0}$ as
\begin{equation}
\begin{split}
u^{\lambda}_{{\bf 0},\theta}\equiv\theta d^{0\lambda}_{\theta},\; \futomoji{\nu}^{\lambda}_{{\bf 0}}&=-\futomoji{u}^{\lambda}_{{\bf 0}}, \;\omega_{{\bf 0}}=0,\\
u^{A}_{{\bf 0},\theta}\equiv d^{0A}_{\theta},\; u^{B}_{{\bf 0},\theta}&=0,\; \futomoji{\nu}^{\lambda}_{{\bf 0}}=-\futomoji{u}^{\lambda}_{{\bf 0}}, \;\omega_{{\bf 0}}=0,\\
u^{B}_{{\bf 0},\theta}\equiv d^{0B}_{\theta},\; u^{A}_{{\bf 0},\theta}&=0,\; \futomoji{\nu}^{\lambda}_{{\bf 0}}=-\futomoji{u}^{\lambda}_{{\bf 0}}, \;\omega_{{\bf 0}}=0.
\end{split}
\end{equation}
The first one is independent of the other two as far as the U(1) symmetry is broken.  
What we do next is to expand the equation around ${\bf k} \approx {\bf 0}$ to find the lowest-energy solution. In order to do this, we expand as 
\begin{equation}
\begin{split}
&u^{\lambda}_{{\bf k},\theta}=u^{(0)\lambda}_{\theta}+u^{(1)\lambda}_{{\bf k},\theta}+u^{(2)\lambda}_{{\bf k},\theta}+\cdots ,\\
&\nu^{\lambda}_{{\bf k},\theta}=\nu^{(0)\lambda}_{\theta}+\nu^{(1)\lambda}_{{\bf k},\theta}+\nu^{(2)\lambda}_{{\bf k},\theta}\cdots,\\
&\omega_{{\bf k}}=\omega^{(0)}+\omega^{(1)}_{{\bf k}}+\omega^{(2)}_{{\bf k}}\cdots ,
\end{split}
\end{equation}
and 
\begin{equation}
\Upsilon({\bf k})=\Upsilon_0+\Upsilon^{(2)}({\bf k})+\cdots.
\end{equation}
In the expansion for $\Upsilon({\bf k})$, there is no first order component of $k$ ($\Upsilon^{(1)}({\bf k})$), because the ${\bf k}$ dependence arises through $\gamma({\bf k})$.
We start with 
\begin{equation}
u^{\lambda}_{{\bf 0},\theta}\equiv \left(\theta+\frac{\partial \omega_{\lambda}}{\partial h}\right) d^{0\lambda}_{\theta},\; \nu^{\lambda}_{{\bf 0}}=-u^{\lambda}_{{\bf 0}}, \;\omega_{{\bf 0}}=0.\label{eq:start1}
\end{equation}
The first-order equation is 
\begin{equation}
\omega^{(1)}_{{\bf k}}\begin{bmatrix}
\futomoji{u}^{(0)}\\
\futomoji{\nu}^{(0)}
\end{bmatrix}=
\Upsilon_0 \begin{bmatrix}
\futomoji{u}^{(1)}\\
\futomoji{\nu}^{(1)}
\end{bmatrix},\label{eq:start}
\end{equation}
where $\futomoji{u}^{(0)}=((\futomoji{u}^{(0)A})^T,(\futomoji{u}^{(0)B})^T)^T$ and $\futomoji{\nu}^{(0)}=((\futomoji{\nu}^{(0)A})^T,(\futomoji{\nu}^{(0)B})^T)^T$.  
With the identity Eq.(\ref{eq:key}), the first-order solution is 
\begin{equation}
u^{(1)\lambda}_{{\bf k},\theta}\equiv \omega^{(1)}_{{\bf k}} \frac{\partial d^{0\lambda}_{\theta}}{\partial h},\; \futomoji{\nu}^{\lambda}_{{\bf 0}}=\futomoji{u}^{\lambda}_{{\bf 0}}.\label{eq:start3}
\end{equation}
We note here that $( \frac{\partial d^{0A}_{1}}{\partial h},\frac{\partial d^{0A}_{0}}{\partial h},\frac{\partial d^{0A}_{-1}}{\partial h},\frac{\partial d^{0B}_{1}}{\partial h},\frac{\partial d^{0B}_{0}}{\partial h},\frac{\partial d^{0B}_{-1}}{\partial h})$ is perpendicular to $(d^{0A}_{1},d^{0A}_{0},d^{0A}_{-1},0,0,0)$ and $(0,0,0,d^{0B}_{1},d^{0B}_{0},d^{0B}_{-1})$. These relations (Eq.(\ref{eq:start1})$\sim$Eq.(\ref{eq:start3}) and the orthogonality) justify starting from Eq.(\ref{eq:start1}).

The second-order equation is 
\begin{equation}
\omega^{(1)}_{{\bf k}}\begin{bmatrix}
\futomoji{u}^{(1)}\\
\futomoji{\nu}^{(1)}
\end{bmatrix}+
\omega^{(2)}_{{\bf k}}\begin{bmatrix}
\futomoji{u}^{(0)}\\
\futomoji{\nu}^{(0)}
\end{bmatrix}=
\Upsilon_0 \begin{bmatrix}
\futomoji{u}^{(2)}\\
\futomoji{\nu}^{(2)}
\end{bmatrix}+
\Upsilon^{(2)}({\bf k}) \begin{bmatrix}
\futomoji{u}^{(0)}\\
\futomoji{\nu}^{(0)}
\end{bmatrix}.
\end{equation}
If we multiply $((\futomoji{u}^{(0)})^T,-(\futomoji{\nu}^{(0)})^T)$ and make use of some property of $W,V$, then we obtain
\begin{equation}
\begin{split}
&(\omega^{(1)}_{{\bf k}})^2\left[\frac{\partial \delta n_A}{\partial h}+\frac{\partial \delta n_B}{\partial h}\right]=4k^2t\phi_A\phi_B\\
&\Longrightarrow \omega_{{\bf k}}^{(1)}=\Bigl(\frac{2t}{\kappa}\phi_A \phi_B\Bigl)^{1/2}k.
\end{split}
\end{equation} 
Here $\kappa=(\frac{\partial \delta n_A}{\partial h}+\frac{\partial \delta n_B}{\partial h})/2$ is the spin susceptibility, which corresponds to the compressibility in bose language.

\end{document}